\documentclass[12pt,fleqn]{article}
\usepackage{times}
\usepackage{graphics,latexsym,psfrag}
\usepackage{amsmath}
\usepackage{amssymb}
\usepackage[pdftex]{graphicx}
\usepackage{epstopdf}
\usepackage{lscape}
\usepackage{color}
\usepackage{subfigure}

\newcommand{\ba}{\begin{array}}
\newcommand{\ea}{\end{array}}

\begin{document}
\newcommand{\be}{\begin{equation}}
\newcommand{\ee}{\end{equation}}
\newcommand{\bc}{\begin{center}}
\newcommand{\ec}{\end{center}}
\newcommand{\bdm}{\begin{displaymath}}
\newcommand{\edm}{\end{displaymath}}
\newcommand{\ds}{\displaystyle}
\newcommand{\p}{\partial}
\newcommand{\INT}{\int\limits}
\newcommand{\SUM}{\sum\limits}
\newcommand{\bfm}[1]{\mbox{\boldmath $ #1 $}}
\renewcommand{\theequation}{\arabic{section}.\arabic{equation}}

\title{\bf Effect of flow on ATP/ADP concentration\\
at the endothelial cell surface:
interplay between \\ shear stress and mass transport }

\author{{\em  Ezio Di Costanzo$^{a}$,   Abdul I. Barakat$^{b}$,
Giuseppe Pontrelli$^{a}\footnote{Corresponding author, Email: giuseppe.pontrelli@gmail.com}$}  \\ \\
$^{a}$Istituto per le Applicazioni del Calcolo - CNR \\
Via dei Taurini 19 -- 00185 Rome, Italy 
\medskip \\
$^{b}$Hydrodynamics Laboratory – CNRS UMR7646 \\
Ecole Polytechnique --  91128 Palaiseau, France
\vspace{4mm}
}


\maketitle
\begin{abstract}
The nucleotides ATP and ADP regulate many aspects of endothelial 
cell (EC) biology, including intracellular calcium concentrations, 
focal adhesion activation, cytoskeletal organization, and cellular motility.
{\em In vivo}, ECs are constantly under flow, and the concentration 
of ATP/ADP on the EC surface is determined by the combined effects of nucleotide convective and diffusive transport as well as hydrolysis by ectonucleotidases on the EC surface. 
In addition, experiments have demonstrated 
that flow induces ATP release from the cells. Previously 
computational models have incorporated the above effects and thus 
described concentration at the EC surface. 
However, it remains unclear what physical processes are responsible 
for nucleotide  regulation. While some EC responses to flow 
have been shown to be directly driven by shear stress, others appear 
to also involve a non-negligible contribution of transport. 
In the present work, 
we develop a mathematical model and perform numerical simulations to investigate 
the relative contributions of shear stress and transport to nucleotide 
concentration at the EC surface, with the effect of cell density.
Because {\em in vitro} experiments are performed
by using confluent cells in some cases and subconfluent cells in other cases,  we also investigate the effect of cell density on the results. 
 The outcomes  of the simulations demonstrate  a complex interplay between shear stress and transport such that transport has 
a significant contribution at certain shear stress values but not at others. 
The effect of transport on nucleotide concentration increases 
with cell density. The present findings enhance our understanding 
of the mechanisms that govern the regulation of such molecules 
at the EC surface under flow. The implications of these findings for downstream responses such as cellular motility merit future investigation. 
\end{abstract}

\section{Introduction}

The extracellular action of the adenine nucleotides adenosine triphosphate (ATP) and adenosine diphosphate (ADP) modulates important processes in vascular endothelial cells (ECs), including the production of vasoactive agents and of intracellular calcium \cite{davies,kac}. Therefore, elucidating the factors that regulate  ATP and ADP concentration at the EC surface is of importance. ATP/ADP  concentration at the EC surface is determined by the balance among three concurrent processes \cite{david,pla1,jon,kim,nollert}: {\em i)} nucleotide convective and diffusive transport to and from the EC surface, {\em ii)} hydrolysis by cell-surface ectonucleotidases that catalyze the reaction sequence ATP $\rightarrow$ ADP $\rightarrow$ AMP, and {\em iii)} ATP release from ECs due to the fluid dynamic shear stress exerted on the cell surface by the flow of viscous blood.

Previous studies have formulated mathematical models that describe the interplay among the three physico-chemical processes described above in various geometries including parallel plate flow chambers \cite{jon,choi}, a channel with a backward facing step \cite{cho}, and in models of arterial segments \cite{come}. These models allow computing the ATP and ADP concentration at the EC surface by solving the advection-diffusion equation within the fluid subject to sink and source terms at the EC surface that respectively represent nucleotide hydrolysis and shear stress-induced ATP release. The overall conclusions from these studies are that flow-induced ATP release has a pronounced impact on the ATP/ADP release at the EC surface and that flow disturbances as would occur in flow separation and recirculation zones is typically associated with significantly reduced concentration.

An important question in EC flow-mediated mechanotransduction is whether a specific biological response to flow is driven directly by the shear force due to the flow, i.e. a shear stress effect, or rather indirectly by flow-mediated alterations in the transport rate of agonists to and from the EC surface, i.e. a shear rate effect. Since shear stress is proportional to the product of the shear rate and the dynamic viscosity of the fluid, one approach to distinguish a shear stress-driven response from a shear rate-driven response is to vary the fluid viscosity while maintaining the same flow rate and to subsequently monitor the biological response of interest. If the biological response is the same despite the viscosity differences, then the response is shear rate-driven (or transport-driven), whereas if the response is different for the different viscosities, then it is shear stress-driven (or force-driven). This approach has been used to demonstrate, for instance, that the activation of flow-sensitive  Cl$^{-}$ channels is driven directly by shear stress \cite{gau}. In the context of the present interest in ATP/ADP concentration at the EC surface, shear stress drives flow-induced ATP release while shear rate determines the rate of nucleotide transport; therefore, both shear stress and shear rate are expected to be important. The primary goal of the present study is to establish when the ATP/ADP response is determined principally by shear stress and when it is driven mainly by shear rate. Because the relative contributions of shear stress and shear rate are expected to depend on EC density, a secondary goal is to elucidate how differences in cell confluence levels modulate the interplay between shear stress and shear rate. \par
The article is organized as follows. In Section 2, we present the model equations and boundary conditions that govern the dynamics of ATP and ADP concentrations within a flow chamber containing ECs. In section 3 we validate the model and, through extensive simulations, we show the sensitivity of the system to the various parameter values including the inter-cell spacing and the relative contributions of shear stress and transport to nucleotide concentration at the EC surface. Finally, section 4 provides general conclusions and some ideas for future studies.

\section{Formulation of the problem and solution methodology}
\setcounter{equation}{0}
Many {\em in vitro} experiments on ECs are conducted under fully developed steady flow conditions in parallel plate flow chambers where the flow field is well characterized. To reproduce such a situation in the current model, we consider a plane 2D channel of height $h$  with a solid (impermeable) cell-free top plate and the ECs located on the bottom plate (Fig. 1 top).  As in previous studies \cite{jon,choi,cho}, a steady fully developed Poiseuille flow is imposed at the channel entrance, with:  

\be
v(y)= 6 \bar v  {y \over h} \left( 1 - {y \over h} \right)
\ee
where  $\bar v$ is the mean longitudinal velocity and $\mu$ the fluid 
dynamic viscosity.
The wall shear stress is given by:
\be
\tau_w= \left. \mu { \p v  \over \p y} \right|_{y=0} ={ 6 \mu  \bar v \over h}
\label{eq1}
\ee

Under the assumption of a dilute solution, the fluid flow is not 
influenced by the solute concentrations. 
Prior to the onset of flow, the fluid within the flow
chamber is assumed to contain cell culture medium with
a uniform ATP concentration $c_0$ taken to be $0.1 \mu M$ in
all simulations. This value corresponds to the
smallest value used in experiments that have reported
flow-induced oscillations in intracellular calcium in the
presence of ATP.  The initial ADP concentration
is taken to be zero in all the simulations. At
$t=0$, cell culture medium with the same ATP concentration
$c_0$ as in the flow chamber is made to flow into the
chamber. Contact between the flowing fluid and the ECs
leads to nucleotide hydrolysis at the cell surface as well
as to flow-induced  ATP release from the ECs. The ATP and ADP
concentration within the flow chamber can be described
by the convection-diffusion equation given by
\be 
{ \p c \over \p t} + v(y) { \p c  \over \p x} = D \left( {\p^2 c \over \p x^2} +
 {\p^2 c \over \p y^2} \right)  \label{eq1b}
\ee 
where $c$ is the nucleotide concentration and $D$ is its
diffusion coefficient in the fluid. Equation
(\ref{eq1b})  applies for both ATP and ADP except that the
values of $D$ are different for the two molecules. 

At the upper wall we impose a zero mass flux condition for both ATP
and ADP; thus
\be
{\p c \over \p y }=0 \qquad\qquad   \mbox{at} \; y=h 
\ee

At the lower wall,  where the ECs are present, the net ATP mass
flux is determined by the rate of ATP hydrolysis by
ecto-ATPases on the cell surface and the rate of shear stress-induced
ATP release by the ECs. Similar to Shen et al. \cite{shen1,shen2},
we assume that the kinetics of ATP hydrolysis are described
by an irreversible Michaelis-Menten formulation,
while flow-induced ATP release due to flow is included as a
separate source term. Thus ATP flux at the EC surface, after changing the sign, is
given as
\be
D_{ATP} {\p c_{ATP} \over \p y }=  { V_{max,T} \; c_{ATP} \over K_T  +c_{ATP}  }
 -S_{T}(\tau_w)    \qquad\qquad   \mbox{at} \; y=0   \label{eq4}
\ee
where $D_{ATP}$ is the diffusion coefficient for ATP in cell
culture medium, $V_{max,T}$  is the maximum enzyme reaction
velocity for ATP hydrolysis, $K_T$ is the Michaelis
constant for the enzyme, and $S_T$ is the source term for
endothelial flow-induced ATP release which, in the most
general sense, is a function of the wall shear stress.
As has been done in previous studies \cite{shen1}, we assume a sigmoidal dependence of ATP release on shear stress  such as:

\be
S_{T}(\tau_w) =S_{max} \left[ 1- \exp \left( \ds{-\tau_w \over \tau_0}\right) \right]^3 
 \quad\qquad  \tau_0= 10 \, \textnormal{dyn cm$^{-2}$ }   
\quad\qquad  S_{max}= 10^{-9} \textnormal{mol m$^{-2}$ s$^{-1}$} \label{eq2}
\ee
(see fig. 2). This profile is consistent with experimental measurements on shear stress-induced ATP release from ECs \cite{bod}. 
At low nucleotide concentrations ($c \ll K_T$), and Eq.  (\ref{eq4}) reduces to
\be
D_{ATP} {\p c_{ATP} \over \p y }=  { V_{max,T} \; c_{ATP} \over K_T    }
 -S_{T}(\tau_w)   \qquad\qquad   \mbox{at} \; y=0  \label{eq5}
\ee

For ADP, the mass flux at $y=0$ is determined by the rate
of ADP degradation at the EC surface and the rate of
ADP production due to ATP hydrolysis. Assuming
Michaelis-Menten kinetics and low  concentrations
as in the case of ATP, this can be formulated as
\be
D_{ADP} {\p c_{ADP} \over \p y }=  { V_{max,D}  c_{ADP} \over K_D}
 - { V_{max,T}  c_{ATP} \over K_T }   \qquad\qquad   \mbox{at} \; y=0 \label{eq6}
\ee
where the ADP-subscripted variables are the ADP
equivalents of those defined above for ATP. Equation (\ref{eq1b})
written for both ATP and ADP and the boundary conditions
described above fully specify the mathematical
model for ATP and ADP concentration within the parallel
plate flow chamber as a function of both space and time.
Table 1 provides numerical values for all the parameters in the model.

\bigskip

To complete specification of the problem, initial concentrations
as well as boundary conditions at
the flow chamber inlet  need to be specified for both
ATP and ADP.  At $t=0$, the ATP concentration is $c_0$ ,
while the ADP concentration is zero throughout the flow
chamber:
\be
c_{ATP}(x,0)= c_0 \qquad\qquad    c_{ADP}(x,0)= 0    
\ee

We wish to model a very long flow chamber while maintaining a high (cell-scale) spatial resolution. To this end,
we impose periodic boundary conditions,  i.e.
\be
c_{ATP}(0,t)=c_{ATP}(L,t)  \qquad\qquad    c_{ADP}(0,t)=c_{ADP}(L,t)   
\ee
 mimicking a closed-loop channel.  This physically corresponds to the typical experimental setup which uses a flow loop with a recirculating fluid (Fig. 1 bottom). 
In this way a relatively small $L$ ($L$ is arbitrary indeed) 
can be considered and a fully
developed flow is automatically sustained.  
Consequently, no inflow/outflow  conditions  on concentration are needed  but,
to  account for a possible ATP supply  coming from the recirculating fluid, 
we add an ATP source $H_t (c_0-c)$ in eqn. (\ref{eq1b}):

\be
{\p c_{ATP} \over \p t} + v(y) {\p c_{ATP} \over \p x}= D_t \left( {\p c_{ATP}^2 \over \p x^2} +
{\p c_{ATP}^2 \over \p y^2}     \right) +  H_t (c_0-c_{ATP})  
\ee
where $H_t$ ($s^{-1}$) is a source rate,   
$H_t(\bar v) = k_t \; \bar v $ . 
The source (and sink) terms come from the external loop, where there is a loss  (at $x=L$) and 
a source $c_0$ (at $x=0$) of concentrations.
$H_t$ is the rate at which ATP crosses the flow chamber
 per unit time, and it
depends on the fluid velocity and the length of the chamber. It replaces the condition $c=c_0$ at the inlet. \par
Similarly, for ADP we have:

\be
{\p c_{ADP} \over \p t} + v(y) {\p c_{ADP} \over \p x}= D_d \left( {\p c_{ADP}^2 \over \p x^2} +
{\p c_{ADP}^2 \over \p y^2}     \right) -  H_d c_{ADP}  \ee
with  $H_d$ ($s^{-1}$) a sink rate, $H_d(\bar v) = k_d \; \bar v $.
The sink term acts as a degradation (or wash out) of ADP, due to the external loop.  As detailed below, this different modeling approach was validated 
against  the results reported in \cite{jon} and was
show to provide excellent agreement.  \par	

In all the simulations, we consider a parallel plate flow chamber of length $L=0.036$ cm  and height $h=0.025$ cm (a typical size of chambers used in experiments). The bottom wall is covered by ECs separated by a distance $d$ apart (Fig. 1 top). The case of confluent cells is recovered by taking $d=0$. The model equations are solved numerically using the finite difference scheme. The computer code developed for this purpose was based on a two-stage corrected Euler formulation with an upwind difference approximation for all spatial derivatives. Near the EC's, a 3-node scheme is adopted to guarantee increased accuracy. We discretize the domain $[0,L]\times[0,h]$ using a uniform mesh with steps $\Delta x=\Delta y=5\times 10^{-4}$ cm, which provides a resolution  of 6 grid nodes per EC (each EC is assumed to have a constant length of $30 \mu m$; see Table 1).

\section{ Results and discussion}
\setcounter{equation}{0}

\subsection{\em Model Validation}
To validate the model and ascertain that the closed loop idealization 
depicted in Fig. 1 provides accurate results, 
we consider the basic case of confluent cells ($d=0$) with no shear 
stress-induced ATP release and simulate three different levels of wall 
shear stress: 0.1, 1, and $10 \, dyne/cm^2$ as has been done in previous 
work \cite{jon}. Figure 3 illustrates the ATP and ADP concentrations 
(normalized by $c_0$) as a function of the transverse coordinate $y$ for
 the three values of wall shear stress. As expected, the ATP concentration 
is lowest at the EC surface ($y=0$) where ATP hydrolysis occurs and increases 
progressively with y. The ADP concentration is largest at the EC surface 
and decreases progressively with $y$. It should be noted that these profiles 
are the same for all values of $x$ because of the infinite length assumption 
implied by the idealized closed loop model. The results of Fig. 3 are
 in excellent agreement with previous results \cite{jon}, thus providing validation for the current modeling approach.  

\subsection{ \em Demonstration of the dependence of ATP/ADP 
concentration on shear rate} 
A primary goal of the present work is to explore the interplay between shear stress  and mass transport (or shear rate)  in determining the ATP/ADP concentration at the EC surface.  As already mentioned, the nucleotide concentration at the EC surface is determined by the combined effects of convective and diffusive transport, hydrolysis, and shear stress-induced ATP release. Since shear stress 
is proportional to the product of shear rate and dynamic viscosity (see eqn. 
(\ref{eq1})), a combination 
of high viscosity and low shear rate can provide the same shear stress as 
a low viscosity and high shear rate.  In the case of ATP, at large shear stress values we expect the effect of release to dominate the effects of transport/hydrolysis, so that the high viscosity/low shear rate combination and the low viscosity/high shear rate combination would yield fairly similar ATP concentrations at the EC surface. On the other hand, at low shear stresses ATP release is limited, and the effects of transport/hydrolysis become more pronounced, leading to significantly lower ATP concentrations at the EC surface for the high viscosity/low shear rate combination where nucleotide residence time near the cell surface is large than for the low viscosity/high shear rate combination where residence time is small. In the case of ADP, there is no release, so we simply expect the effects of transport/hydrolysis to be more pronounced for the high viscosity/low shear rate combination than for the low viscosity/high shear rate combination for all values of shear stress, although the difference between these two combinations will depend in a complex manner on shear stress level since ADP and ATP concentrations are coupled (ATP hydrolysis is the source of ADP). 
We now wish to demonstrate 
these notions in a series of simulations.  \par

An example of how shear rate can modulate ATP/ADP concentration 
independently of shear stress is shown in  Fig. 4, where simulations
 were performed for the same value of wall shear stress ($0.1 dyne/cm^2$) 
and two different shear rates by varying the viscosity. The simulations 
were performed for the case of a semi-confluent endothelial layer ($d=1$) 
and for the shear stress-induced ATP release profile shown in Fig. 2. 
The fact that the ATP and ADP profiles within the flow chamber are
 different at the two different shear rates despite the same shear 
stress value shows that transport/hydrolysis is the primary driver of nucleotide 
concentration in this case. 

\subsection{\em Demonstration of the interplay between shear stress and shear rate}
We have carried out a series of simulations to explore the interplay 
between shear stress and shear rate in regulating ATP/ADP concentration 
at the EC surface. Because we expect cell density to play a role in the 
interplay between shear stress and shear rate, we have complemented 
the simulations on confluent EC monolayers with simulations for three 
cases of sub-confluent cells where the spacing between cells 
($d$ in Fig. 1 top) takes on the value of 0.5, 1, or 2 cell lengths. 
Thus, the four series of simulations have the alternate sequence of cells 
and cell-free spaces $ ...EC-d - EC - d....$, $d$ varied  as: \\
\\
Case EE: $d=0$ (confluent cells) \\ 
\\
Case Ed/2: $d=0.5$ EC \\
\\
Case Ed: $d=$EC \\
\\
Case E2d: $d=2$ EC \\

For each of the above four cases, the dependence of ATP and ADP concentration 
on shear stress and shear rate has been studied assuming the sigmoidal shear 
stress-induced ATP release depicted in Fig. 2. These time-dependent 
simulations are run to steady-state using a discretization time step 
in the range ($6\times 10^{-5}  -  4\times 10^{-3}$) $s$ in order to satisfy 
the CFL condition for each value of $\tau_w$.  \par
Figure 5 depicts the time evolution of the mean normalized concentrations 
of ATP and ADP in the entire flow channel for wall shear stresses ranging 
from $0.1$ to $40 \, dyne/cm^2$  (shear stress-induced ATP release shown 
in Fig. 2) and for the two extreme cases of EC density: confluent 
cells (case EE, left panels) and sparse cells (case E2d, right panels). 
To delineate shear stress vs. shear rate effects, each shear 
stress level is attained either via a combination of high viscosity and 
low shear rate (top panels) or low viscosity and high shear rate (bottom
 panels). In these simulations, the mean concentrations are computed as:

\be
\bar c(t)= {1 \over L h} \int_0^L \int_0^h c(x,y,t) dx dy \label{yu1}
\ee

The results demonstrate largely similar behavior for confluent and 
subconfluent cells (compare the right and left columns); however, the sensitivity 
of ATP/ADP concentration to shear stress is higher for the confluent cells 
than for the subconfluent cells in all cases. The ATP concentration at the EC surface increases with shear stress; at the lowest shear stress ($0.1 dyne/cm^2$), ATP release is minimal so that ATP hydrolysis leads to a significant drop in ATP concentration; however, as the shear stress increases, ATP release increases thereby compensating (and in some cases overcompensating) for the effect of hydrolysis. 
Comparing the top and bottom 
panels reveals that for any value of shear stress, the ATP concentration is
 higher  when that shear stress is obtained 
via a low viscosity-high shear rate combination (bottom panels) than via a 
high-viscosity-low shear rate combination (top panels). This is attributable to the fact that a higher shear rate allows less time for nucleotide hydrolysis. The ADP concentration follows a more complex behavior: it decreases with shear stress for the low shear rate case (Fig. 5 top panels) while the opposite occurs for the high shear rate case (Fig. 5 bottom panels). This suggests that at low shear rate the ADP source term due ATP hydrolysis exceeds the sink term due to ADP hydrolysis, while the reverse is true at high shear rate. A final observation is that the time needed to attain steady-state 
is shorter for the low viscosity-high shear rate combination (bottom panels) 
than for the high-viscosity-low shear rate combination (top panels). 
This finding is not surprising since the higher transport rates associated 
with higher shear rates accelerate the approach to steady-state. 

To gain a better appreciation for the interplay between shear stress and 
shear rate in the regulation of ATP/ADP at the EC surface, 
Fig. 6 depicts the ratio of the average ATP, ADP  and ATP+ADP concentrations at the EC 
surface for the case of the high viscosity-low shear rate combination to those obtained for the case of low viscosity-high shear rate combination for every shear stress 
value studied ($0.1-40 \,dyne/cm^2$) and for 4 different 
levels of cell density 
ranging from sparse (E2d case) to confluent (EE case), assuming 
the sigmoidal shear stress-induced ATP release profile 
in eqn. (\ref{eq2}). The simulations are performed both in the presence (top row) and absence (bottom row) of shear stress stress-induced ATP release. The average ATP and ADP concentrations at the EC surface  
$\tilde c_w$ are computed as follows:

\be
\tilde c_w(t)= {1 \over \tilde L} \int_{\Omega_c} \tilde c(x,0,t) dx 
\ee
(the $\,\tilde{} \, $ accent indicates a restriction to the EC surface 
$\Omega_c$). 
In fig. 6 a ratio of 1 indicates that  the concentration for the high viscosity-low shear rate combination is identical to that for the low viscosity-high shear rate combination, which indicates that 
transport has no effect (no sensitivity 
to shear rate) and that any observed nucleotide-mediated effect occurs 
via shear stress only. The farther the ratio is from 1, the larger the 
contribution of transport. In all cases, the results demonstrate that the contribution 
of transport to the ATP/ADP concentrations at the EC surface increases with 
cell density; thus, at a given value of wall shear stress, transport has a 
larger contribution to regulating nucleotide cell-surface concentration 
for confluent cells than for sparse cells.

Let us first examine the results when shear stress-induced ATP release is included (top panels). For ATP (Fig. 6, top left), the contribution of transport is largest (i.e. the ratio is farthest from 1) at the lowest shear stress ($0.1 dyne/cm^2$). In that case, ATP release is negligible and the ratio is less than 1, reflecting the increased ATP hydrolysis (and thus lower ATP concentration) at the lower shear rate. As the shear stress increases, ATP release kicks in and progressively dominates the effect of transport (i.e. the ratio increases towards 1). The
contribution of transport disappears at a shear stress of $\sim 8 dyne/cm^2$ (where the ratio is 1) but then increases again at the higher shear stresses as ATP release reaches its maximum and overwhelms ATP removal. This is particularly true at the low shear rates where ATP removal is slowest and thus leads to a low shear rate-to-high shear rate ratio larger than 1.

For ADP (Fig. 6, top center), the ratio is always larger than 1, indicating that transport is always significant and that at any value of shear stress, the ADP concentration at the EC surface is always larger at low shear rate than at high shear rate. This behavior reflects the balance of the ADP source (from ATP hydrolysis), ADP removal from the cell surface by advection/diffusion, and  ADP hydrolysis as explained above for Fig. 5. The effect of transport increases initially with shear stress, reaches a peak, and then begins to decrease. Interestingly, the shear stress level at which the contribution of transport is largest depends on cell density: it is $20 dyne/cm^2$ for a confluent EC monolayer and $2.5 dyne/cm^2$ for the most sparse case considered. 

If we consider the combined nucleotide (ATP+ADP) concentration (Fig. 6, top right), transport has a negligible effect at the lower shear stresses as the ATP and ADP effects counteract one another. The effect of transport becomes increasingly more significant as the shear stress increases, peaks at a shear stress of $20 dyne/cm^2$, and then decreases slowly.

In the absence of ATP release (Fig. 6, bottom panels), transport is important for ATP at the lower shear stress levels, but the effect decreases progressively as shear stress increases. Transport is always an important contributor in the case of ADP, with largely similar behavior as described above when ATP release is present. Interestingly, when we consider ATP+ADP, the sensitivity to transport is small and practically non-existent above a shear stress of $ \sim 5 dyne/cm^2$. 


\section {Conclusions}
The present work aimed to provide an understanding of how flow in arteries regulates the concentrations of the adenine nucleotides ATP and ADP at the EC surface. To this end, we focused on the interplay between shear stress and shear rate in order to establish the contribution of transport. The results demonstrate that this interplay is quite complex, leading to particular shear stress regimes where transport considerations are dominant and others where they are negligible. Because ATP and ADP regulate critical EC flow responses including intracellular calcium levels and cell migration, the present findings provide an appreciation for the contribution of transport to important EC mechanotransduction events.

\section*{Acknowledgments}
We are grateful to Dr. R. Natalini for his valuable discussions and
helpful comments. This work was supported in part by an endowment in Cardiovascular Bioengineering from the AXA Research Fund and in part by the Italian INDAM-GNFM.

\begin{table}[h!]
 \footnotesize
 \begin{center}
\begin{tabular}{|c|c|}
\hline
 Parameter  & Value (ATP  $-$  ADP)    \\
\hline
\hline
$D  (\rm{cm^2  s^{-1}})$   & $2.36 \cdot 10^{-6}   -  2.57 \cdot 10^{-6}$    \\
 \hline
$K_m ( \mu M=10^{-6}\rm{mol \, l^{-1}} )$   & $475  \quad - \quad  155  $     \\
 \hline
$V_{max} \rm{(mol \, s^{-1} \, cm^{-2})}$  &  $  0.8 \cdot 10^{-10} \quad - \quad 0.1  \cdot 10^{-10}  $ \\
 \hline
 $\mu \rm{(N \, s \, m^{-2})}$  &   $7.77 \cdot 10^{-4}  (7.77 \cdot 10^{-3} g 
\, cm^{-1} \, s^{-1} )$  \\
 \hline
$L(\rm{cm})$ &  0.036  \\
 \hline
$h(\rm{cm})$ & 0.025 \\
 \hline
$\tau_w(\rm{dyn \, cm^{-2}})$  & various values \\
 \hline
$\tau_0(\rm{dyn \, cm^{-2}})$  & 10\\
 \hline
$k_t(\rm{cm^{-1}})$  & $0.35$ \\
 \hline
$k_d(\rm{cm^{-1}})$  & $0.35$ \\
 \hline
$S_{max}(\rm{mol\, m^{-2} \, s^{-1}})$  & $10^{-9}$ \\
 \hline
$c_0(\mu M=10^{-6}\rm{mol \,  l^{-1}} )$  & $10^{-1} - 0$ \\
 \hline
$EC$ (\rm{cm})  & $0.003$  \\
 \hline
$d$(\rm{cm})  & $0 - 0.0015 - 0.003 - 0.006 $  \\
\hline\hline
\end{tabular}
\caption{Table of parameters. All the parameters are taken from Ref [5], with the
exception of $L$, $k_t$ and $k_d$. The first one guarantees
a sensible number of cells and cell-free space sequence, the other two ensure 
sink and source rates consistent with the other parameters in a closed-loop channel.  }
\end{center}
\end{table}

\newpage

\bigskip\bigskip 
\begin{figure}
\centering\scalebox{0.9}{\includegraphics{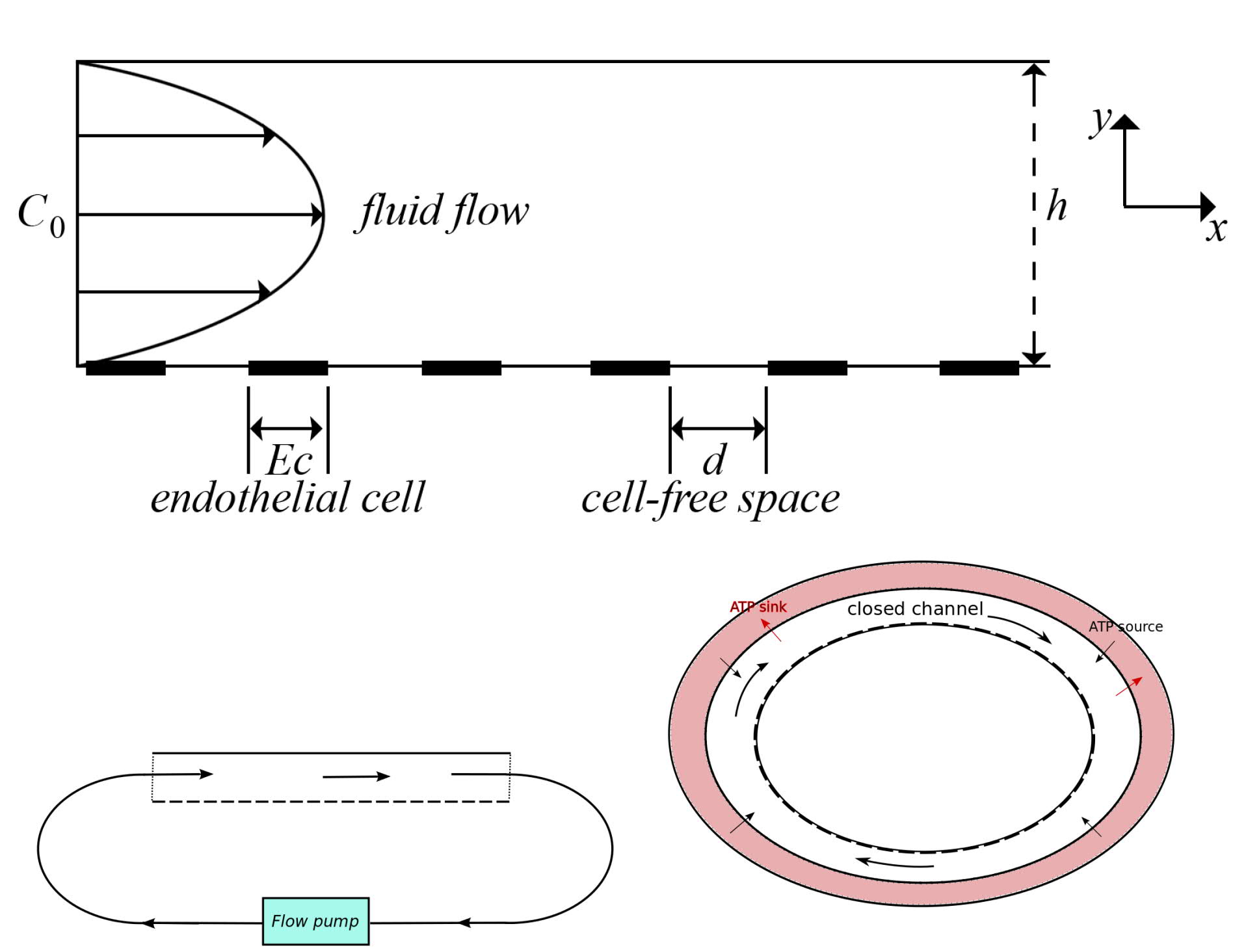}} 
\caption{Schematic representation of the channel experimental
set up (top and bottom left) and its idealization with a closed toroidal shaped tube (bottom right). 
The diffused source (sink) terms in eqns. (3.2)-(3.3) arise from having
 neglected the external circuit and replaced the inlet/outlet conditions.}
\label{channel}
\end{figure}

\bigskip\bigskip 
\begin{figure}
\centering\scalebox{0.9}{\includegraphics{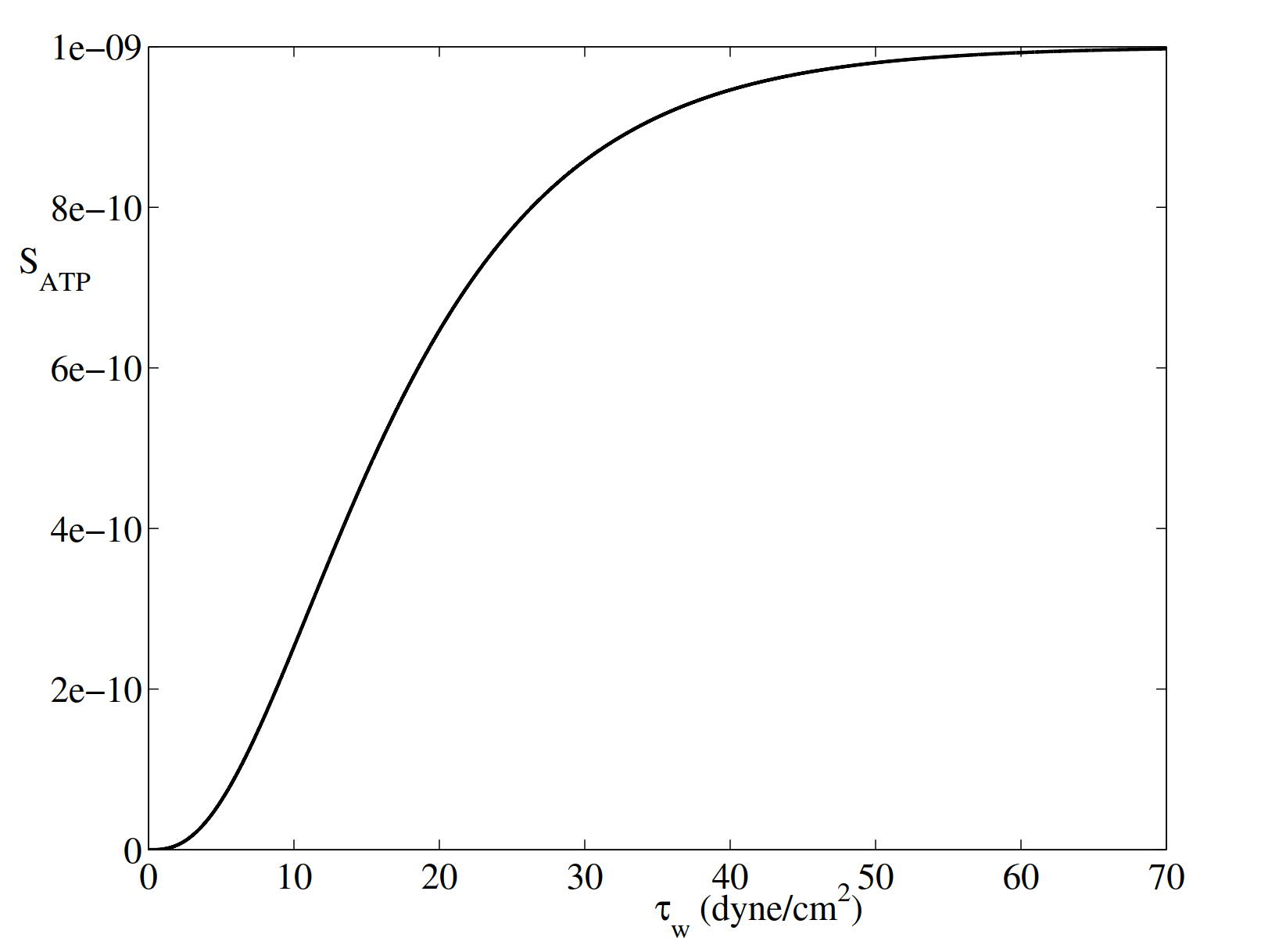}} 
\caption{Shear stress dependent ATP release profile
 as in eqn (\ref{eq2}). A nearly linear
release  between 10 and 25 $dyn \,cm^{-2}$ is accounted \cite{bod}. }
\label{stress}
\end{figure}

\bigskip\bigskip 
\begin{figure}
\centering\scalebox{1}{\includegraphics{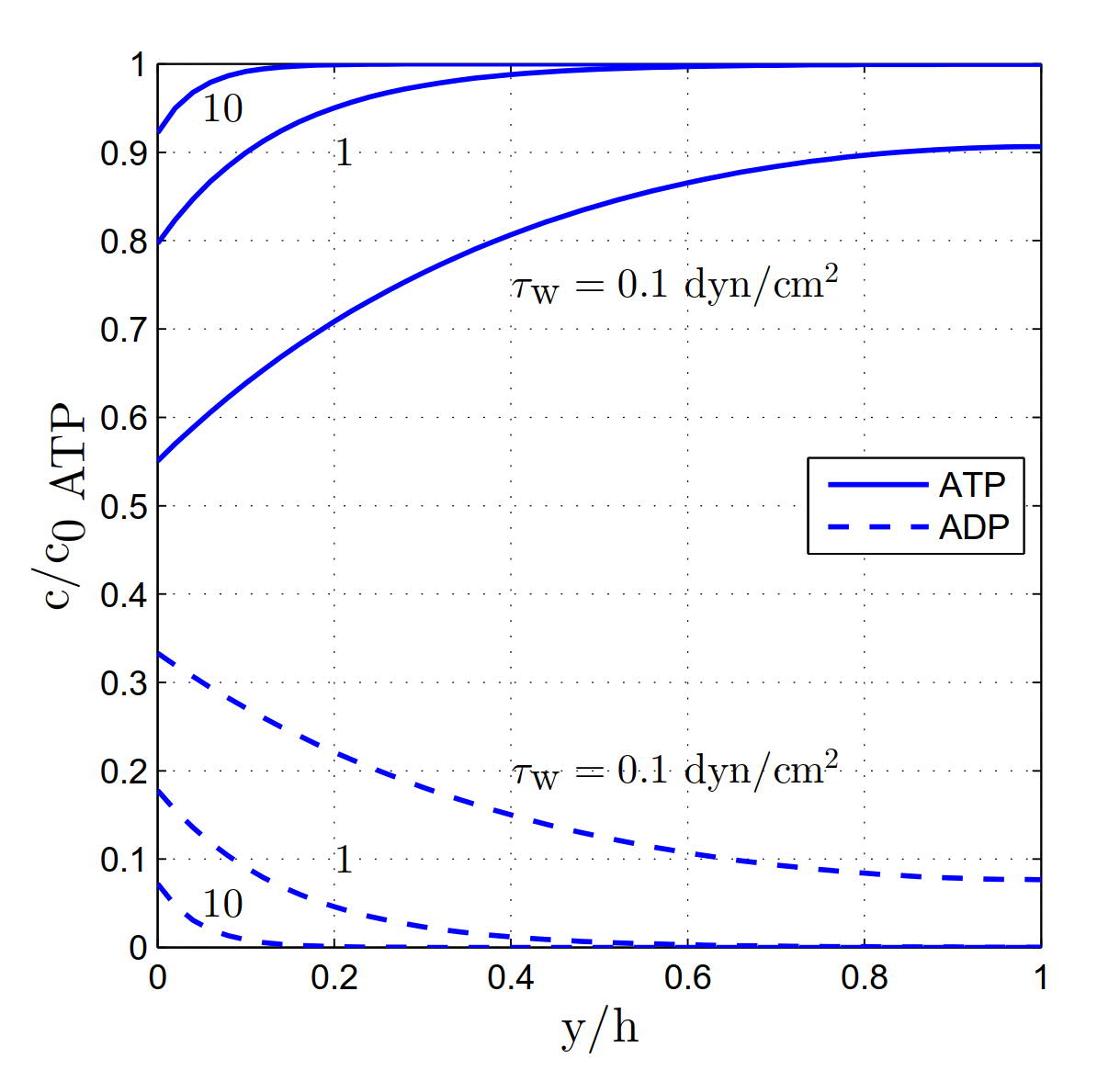}} 
\caption{Steady state ATP (continuous)  and ADP (dashed)  concentration profiles as a function of nondimensional transverse distance, for no flow induced ATP release at $\tau_w=0.1, 1, 10 \, dyn/cm^{2}$. Due to the fully developed flow in the current model, such profiles are independent of the axial coordinate \cite{jon}.} 
\label{stress}
\end{figure}

\begin{figure}
\centering\scalebox{1}{\includegraphics{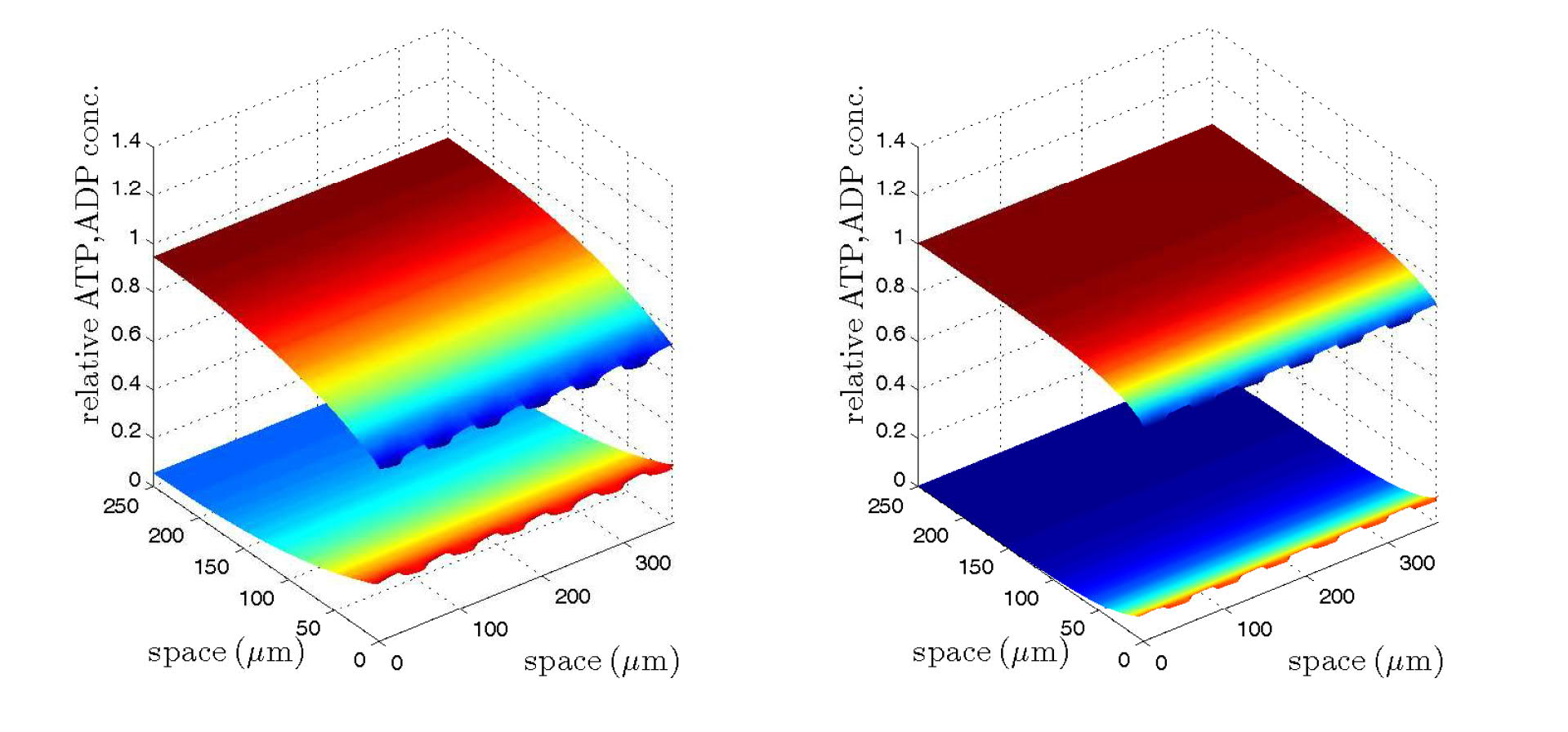}} 
\caption{Influence of shear rate and viscosity on steady state
normalized ATP/ADP concentrations at the same value of  $\tau_w=0.1 dyn/cm^2 $ (Left: $\mu=7.77 \cdot 10^{-3}  g  \,cm^{-1} s^{-1}, \bar v=5.36 \cdot 10^{-2}  cm \, s^{-1}$ , Right:  
$\mu=7.77 \cdot 10^{-4}  g \,cm^{-1} s^{-1}, \bar v=5.36 \cdot 10^{-1} cm \, s^{-1}$).}
\end{figure}

\newpage 
\begin{figure}
\centering\scalebox{1}{\includegraphics{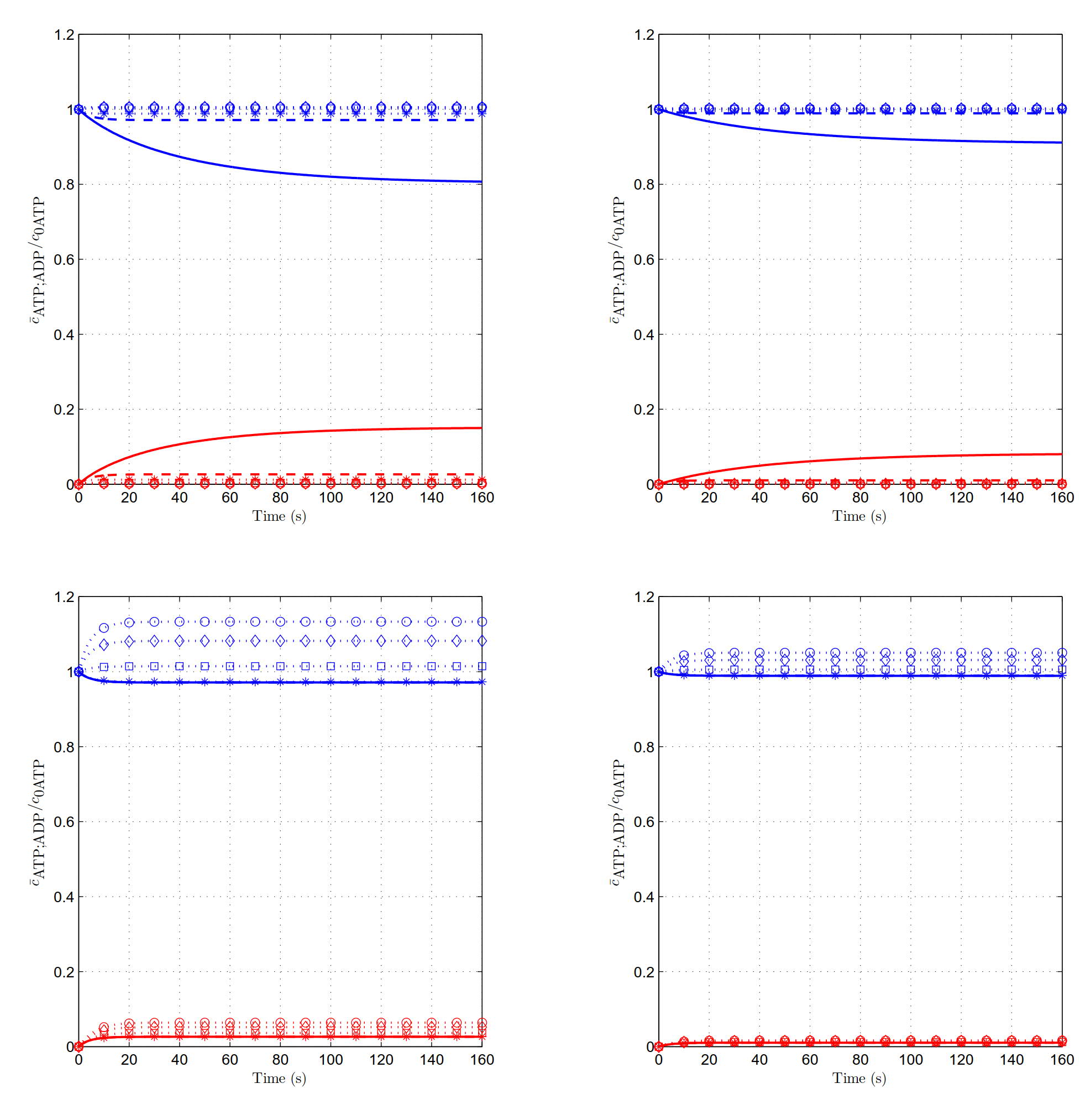}} 
\caption{Influence of shear rate, viscosity and spacing on the values of $\tau_w$  (see eqn. (2.2)).
The figure depicts the mean ATP (blue curve) and ADP (red curve) concentrations $\bar c$
(computed as in eqn (\ref{yu1})) in the channel and times to steady state at six $\tau_w$ ($dyn \, cm^{-2}$) values: $0.1$ (continuous),  $1$ (dashed),  $2.5$ (dot-starred),  $10$ (dot-squared),  $20$ (diamond), $40$ (dot-circled). 
TOP:  fixed $\mu =7.77 \cdot 10^{-3} g \, cm^{-1} \, s^{-1}$, with varying  $\bar v$;   
BOTTOM: fixed $\bar v=5.36 \cdot 10^{-1} cm \, s^{-1}$, with varying $\mu$;
LEFT: spacing EE; RIGHT: spacing E2d.}
\end{figure}

\begin{figure}
\centering\scalebox{1}{\includegraphics{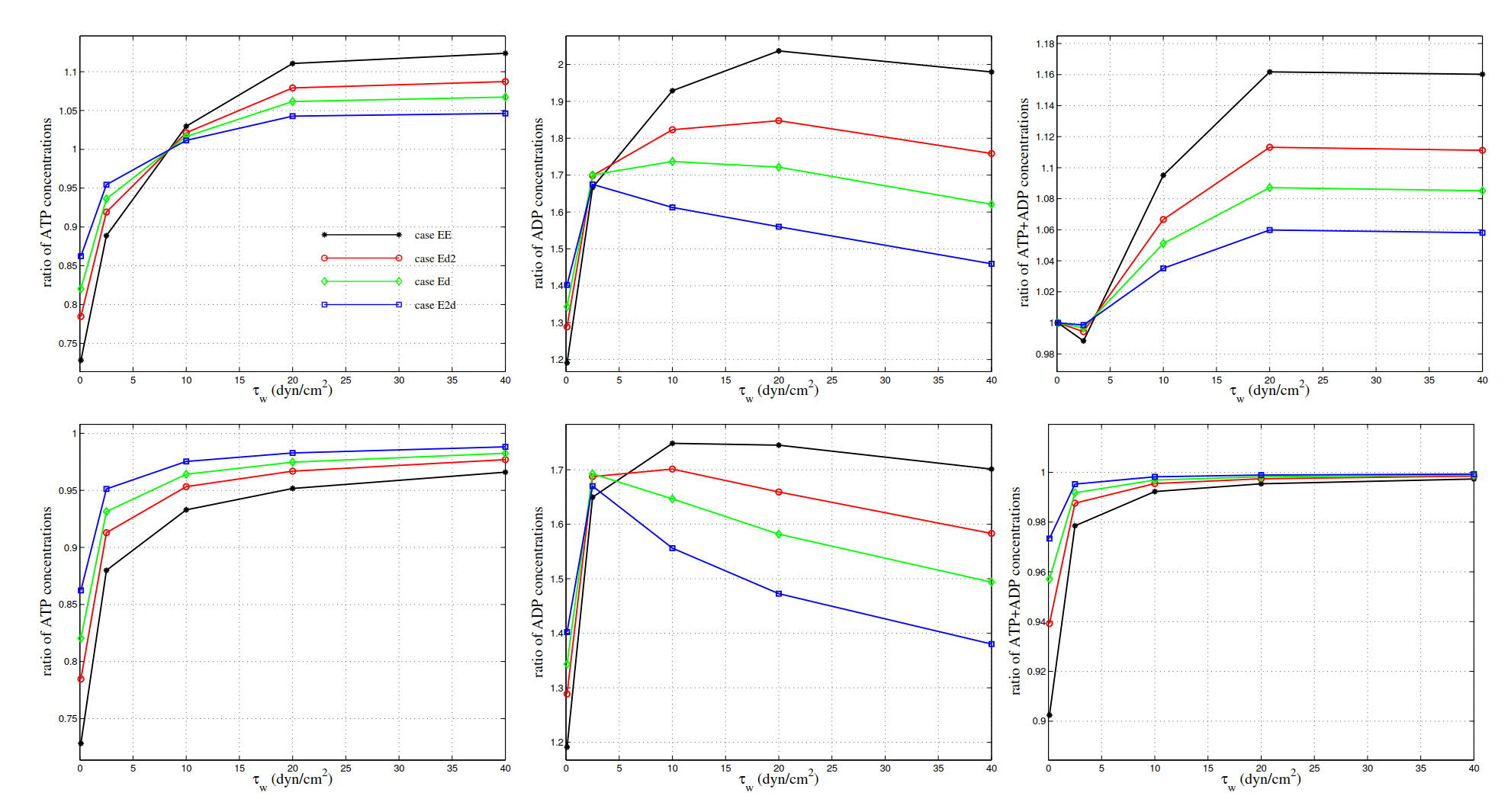}} 
\caption{Sensitivity to shear rate of  ATP (left), ADP (center) and ATP+ADP (
right)  steady state concentration ratio at the EC surface for different $\tau_w$ and at different cell densities (case with release at top, case no release at bottom).}
\end{figure}

\end{document}